\documentclass{article}

\usepackage{arxiv}

\usepackage[utf8]{inputenc} 
\usepackage[T1]{fontenc}    
\usepackage{hyperref}       
\usepackage{url}            
\usepackage{booktabs}       
\usepackage{amsfonts}       
\usepackage{nicefrac}       
\usepackage{microtype}      
\usepackage{lipsum}
\usepackage{graphicx}
\usepackage{siunitx}        
\usepackage{arydshln}       
\usepackage{amsmath}        
\usepackage{xspace}         
\usepackage{xcolor}

\graphicspath{ {./images/}{./assets/} } 

\newcommand{\methodname}{SP-I2I\xspace}
\definecolor{cvprblue}{rgb}{0.21,0.49,0.74}
\hypersetup{
    colorlinks,
    linkcolor=cvprblue,
    citecolor=cvprblue,
    urlcolor=cvprblue
}
\newcommand{\pdfAuthorBlock}{%
    \begin{center}
        \newcommand{\authCell}[3]{%
            \begin{minipage}[t]{##3}
                \centering
                \bfseries ##1 \\
                \normalfont \small ##2
            \end{minipage}%
        }
        \authCell{Aditya Pratap Singh}
                 {Dept. of Computer Science \& Engineering \\ University of Michigan}
                 {0.32\textwidth}
        \hfill
        \authCell{Shrey Shah}
                 {Dept. of Robotics \\ University of Michigan}
                 {0.32\textwidth}
        \hfill
        \authCell{Ramanakumar Sankar}
                 {Space Sciences Laboratory \\ University of California, Berkeley}
                 {0.32\textwidth}

        \par\vspace{0.3in} 

        \authCell{Emma Dahl}
                 {Dept. of Geological \& Planetary Sciences \\ California Institute of Technology}
                 {0.48\textwidth}
        \hfill
        \authCell{Gerald Eichstädt}
                 {Independent Researcher}
                 {0.48\textwidth}

        \par\vspace{0.3in} 

        \authCell{Georgios Georgakis}
                 {Jet Propulsion Laboratory \\ California Institute of Technology}
                 {0.48\textwidth}
        \hfill
        \authCell{Bernadette Bucher}
                 {Dept. of Robotics \\ University of Michigan}
                 {0.48\textwidth}
    \end{center}
}

\author{
  Aditya Pratap Singh, Shrey Shah, Ramanakumar Sankar \\
  Emma Dahl, Gerald Eichstädt \\
  Georgios Georgakis, Bernadette Bucher
}

\title{Structure-Preserving Unpaired Image Translation to Photometrically Calibrate JunoCam with Hubble Data}

\makeatletter
\renewcommand{\@maketitle}{%
  \vbox{%
    \hsize\textwidth
    \linewidth\hsize
    \vskip 0.1in
    \@toptitlebar
    \centering
    {\LARGE\scshape \@title\par}
    \@bottomtitlebar
    \vskip 0.4in
    \pdfAuthorBlock 
    \vskip 0.1in \@minus 0.1in
  }
}
\makeatother

\begin{document}
\maketitle

\maketitle
    
    


\newcommand{\authCell}[3]{%
    \begin{minipage}[t]{#3} 
        \centering
        \bfseries #1 \\
        \normalfont \small #2
    \end{minipage}%
}





\vspace{0.3in} 


\begin{abstract}
Insights into Jupiter's atmospheric dynamics are vital for understanding planetary meteorology and exoplanetary gas giant atmospheres. To study these dynamics, we require high-resolution, photometrically calibrated observations. Over the last 9 years, the Juno spacecraft’s optical camera, JunoCam, has generated a unique dataset with high spatial resolution, wide coverage during perijove passes, and a long baseline. However, JunoCam lacks absolute photometric calibration, hindering quantitative analysis of the Jovian atmosphere. Using observations from the Hubble Space Telescope (HST) as a proxy for a calibrated sensor, we present a novel method for performing unpaired image-to-image translation (I2I) between JunoCam and HST, focusing on addressing the resolution discrepancy between the two sensors. Our structure-preserving I2I method, \methodname, incorporates explicit frequency-space constraints designed to preserve high-frequency features ensuring the retention of fine, small-scale spatial structures – essential for studying Jupiter's atmosphere. We demonstrate that state-of-the-art unpaired image-to-image translation methods are inadequate to address this problem, and, importantly, we show the broader impact of our proposed solution on relevant remote sensing data for the pansharpening task.
\end{abstract}


\section{Introduction}

Jupiter's atmosphere is a complex and highly time variable dynamic system which impacts global phenomena across the planet including ammonia and water cycles and the planet's heat budget. Understanding Jupiter's atmosphere and the associated environmental phenomena provides a gateway to improving understanding of atmospheric circulation on Earth as well as gas giants outside the Solar System. A primary mechanism for heat transfer in giant planet atmospheres is moist convection \cite{Gierasch2000}, which has been shown in part to manifest at small spatial scales (or meso scales) of about 100-1000 km \cite{Orton2021}. The ability to study these meso-scale storms on Jupiter quantitatively is currently hindered by the low spatial resolution of existing photometrically calibrated Earth-based and orbital instruments, such as the Hubble Space Telescope (HST).

\begin{figure}[htbp]
  \centering
  \includegraphics[width=0.8\linewidth]{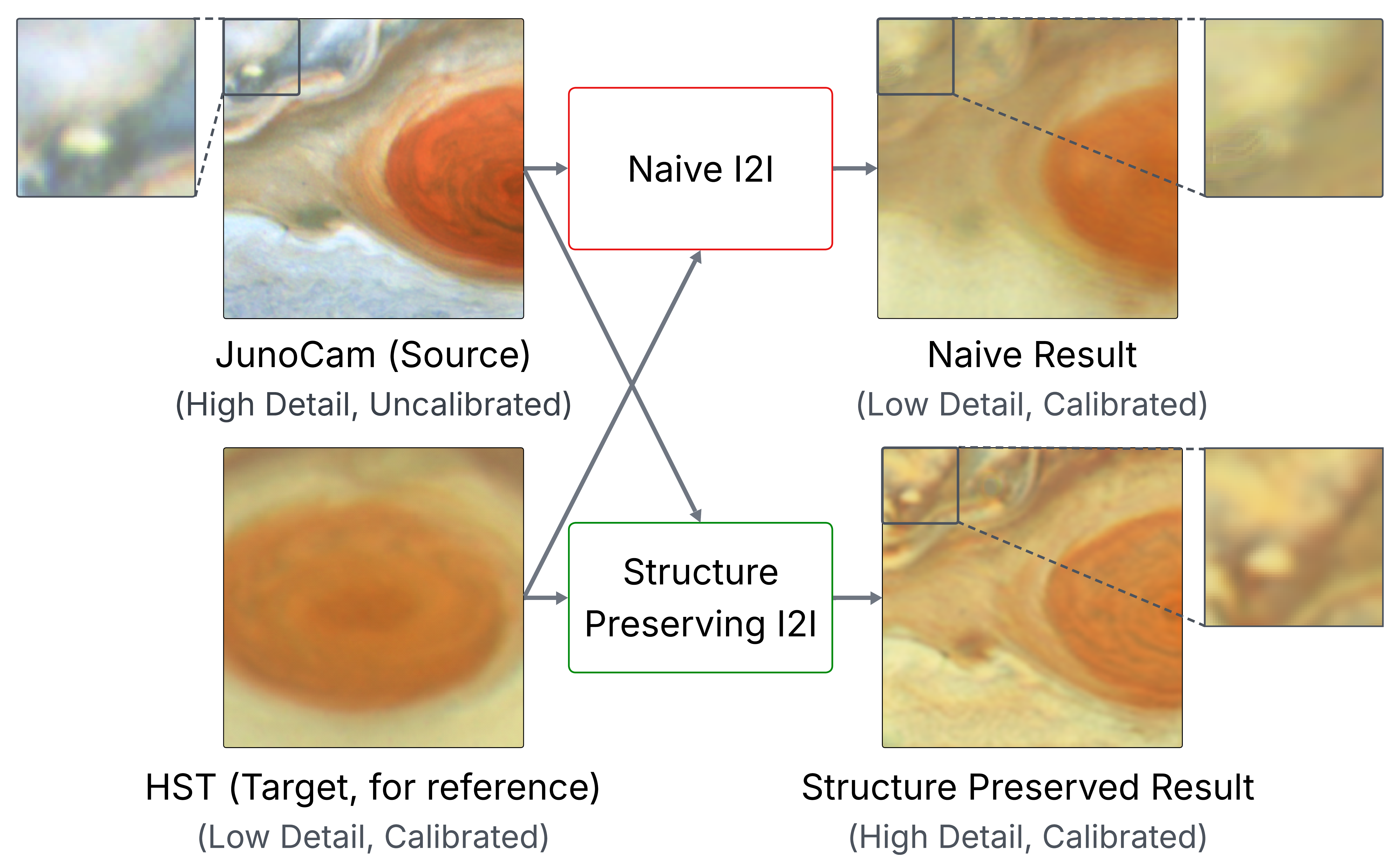}
  \caption{A comparison of naive vs. structure-preserving translation for calibrating high-resolution JunoCam images with low-resolution HST data. While Naive I2I methods lose fine storm details, our Structure-Preserving I2I method successfully retains this critical, high-frequency information.}
  \label{fig:title}
\end{figure}

NASA’s Juno mission was launched to explore Jupiter equipped with JunoCam, a visible light camera which has been operating for the last 9 years. JunoCam is a push-frame imager that captures sequential color filter strips (red, green, and blue) as the spacecraft spins, and is producing observations of meso-scale features in unprecedented detail. However, it was included in the mission primarily for public outreach and lacks the absolute photometric calibration~\cite{Sirianni_2005, diazmiller2006photometricastrometriccalibrationjwst} required to convert its raw sensor readings into physical units of radiance~\cite{Hansen2017}. This renders the spectral information from each filter uncertain, preventing the application of many quantitative scientific analysis techniques needed to characterize cloud properties. A workaround would be to indirectly calibrate JunoCam by learning a linear transformation from observations of an existing calibrated sensor~\cite{Cao2014, Teillet1997}, but no such sensor observations exist that have the necessary overlap with JunoCam.

For the last 10 years, HST has captured annual observations of the giant planets as a part of a long-term monitoring campaign, the Outer Planet Atmospheres Legacy (OPAL)~\cite{wong2025hubble}. These observations include yearly global maps of Jupiter from HST's Wide Field Camera 3 (WFC3) while Juno has been orbiting the planet~\cite{opal1,opal2}, enabling a possible point of comparison between calibrated and uncalibrated datasets. However, JunoCam's and HST's observations are fundamentally unpaired with significantly different viewing geometries and sensor specifications leading HST to have much lower spatial resolution than JunoCam along with variable filter responses that affect the sensor's spectral sensitivities. Asynchronous observations also preclude co-registered views of the same atmospheric features due to the dynamic nature of Jupiter's atmosphere.

Recently, there has been significant progress in the task of unpaired image-to-image translation~\cite{pang2021image,safayani2025unpaired}, which learns to transform an image from a source domain to a target domain while preserving the semantic and structural coherence of the source image. A promising direction is to cast this task as probabilistic interpolation between the source and target distributions by solving a stochastic entropy-regularized optimal transport (OT) problem, known as the Schrödinger Bridge Problem (SBP)~\cite{leonard2014survey,de2021diffusion}. This approach has been successfully adapted for unpaired image-to-image translation~\cite{de2024schrodinger,kim2024unpaired} due to its flexibility, enabling translation between two arbitrary distributions. However, when the target domain has inherently lower resolution (e.g., HST) than the source domain, SBP results in smoothing the translated image, incapable of preserving high-frequency details. Therefore, naively applying SBP for calibrating JunoCam, results in the suppression of meso-scale atmospheric phenomena that are expected to provide scientific value.
 
In this work, we introduce a new method, \methodname, which is an extension to the SBP that preserves high-frequency details in the translated image, and is particularly robust to resolution differences between the two domains -- see Figure~\ref{fig:title}. Using \methodname, we formulate the photometric calibration of JunoCam as an image-to-image translation task, with source and target distributions instantiated as JunoCam and HST observations respectively. Our main idea is to incorporate a Laplacian pyramid \cite{1095851} when solving the optimal transport problem, allowing for decoupled learning of spectral properties, paired with a high-frequency Laplacian consistency loss \cite{bojanowski2019optimizinglatentspacegenerative} to explicitly guide fine grained spatial structure. 

Our experiments on a newly introduced dataset \textit{JunoCam2HST} demonstrate that our method is able to robustly retain source structure compared to existing image-to-image translation methods. Furthermore, we consider the broader impact of our extended SBP formulation and show its application on the pansharpening task, that attempts to translate remote sensing data of varying resolutions.
In summary, our contributions are as follows:
\begin{itemize}
    \item We propose a sensor-agnostic data-driven solution to the problem of photometric calibration of JunoCam that is expected to produce a high quality resource and allow scientists to study Jupiter’s cloud structure with great detail.
    \item We introduce a general extension of the SBP problem that is robust to resolution differences between source and target domain distributions.
    \item We present a new dataset tailored for investigating the challenging conditions for translating JunoCam to HST.
    \item We demonstrate the ability of our method to preserve high-frequency details and its broader usage on remote sensing data. 
\end{itemize}

\section{Related Work}

\noindent\textbf{Image-to-Image Translation.} Image-to-image translation (I2I) aims to generate an image target distribution while keeping structural similarities to the source domain. Paired methods utilize aligned image pairs to learn a direct mapping from source to target images~\cite{pix2pix}. In contrast, unpaired image translation does not require data of aligned image pairs. Foundational work focused on the notion of cycle consistency with the intent of achieving bidirectional translatability to preserve image content \cite{CycleGAN2017,DiscoGan2017, Dualgan2018}. Subsequent work proposed using shared-latent space assumptions, mapping images from different domains to a common representation~\cite{UNIT2018}. To further capture domain-invariant and domain specific properties, subsequent methods disentangled the latent representations~\cite{drit2018, munit2018}. However, this popular cycle consistency design in unpaired image translation models is over-restrictive and causes artifacts, leading to the development of unidirectional translation models \cite{gc_gan2019, cut2020}. Parallel architectural innovations, including attention mechanisms \cite{aggan2019,attentiongan2021}, transformer networks \cite{ittr2022}, deeper hierarchical features \cite{deepi2i2020}, and improved normalization \cite{instaformer2022, pixelhasitsmoment2024} further refined translation quality and control in subsequent I2I methods. 

The recent success of diffusion based methods in image synthesis \cite{ddpm2020, iddpm2021, latentdiffusion2022} sparked an interest in utilizing such methods in I2I tasks. Diffusion based I2I methods achieve translation using guidance mechanisms \cite{palette2022, egsde2022} or by circumventing the Gaussian prior assumption present in diffusion models\cite{bbdm2023}. Diffusion-based approaches are the current state of the art technique for image translation. However, existing methods fail in our setting, where the source domain (JunoCam) contains critical high-frequency details that are absent in the low-resolution target domain (HST) - standard models incorrectly learn to remove these details. Our work extends this direction by incorporating explicit supervision of high-frequency spatial structures while learning an unsupervised translation of color properties.

\noindent\textbf{Frequency Dependent Losses.} Neural networks exhibit an inherent spectral bias, which is a tendency to learn low-frequency components of data more easily than high-frequency ones\cite{rahaman2019spectralbiasneuralnetworks, basri2020frequencybiasneuralnetworks}.  This bias causes models like Variational Autoencoders (VAEs) to lack sharp, high-frequency details, resulting in blurry image generation \cite{björk2022simplerbetterspectralregularization}. This problem is often tackled through explicit constraints in the frequency domain \cite{zhu2021wavelet, bredell2023explicitlyminimizingblurerror}.

In Generative Adversarial Networks (GANs) \cite{goodfellow2014generativeadversarialnetworks}, the frequency domain gap arises due to the discriminator's struggle to provide an informative training signal for high frequency details \cite{schwarz2021frequencybiasgenerativemodels}. This lack of supervision for high frequency details causes the generator to introduce unnatural artifacts in the generated images. To combat these artifacts, numerous methods use frequency separation via high and low-pass filters to explicitly supervise the low and high frequencies \cite{9150978, fritsche2019frequencyseparationrealworldsuperresolution}. Architectural solutions like SSD-GAN \cite{chen2020ssdganmeasuringrealnessspatial} incorporate a frequency-aware classifier to force the generator to match the frequency spectrum of real data. Other approaches modify the training objective ; for instance, Focal Frequency Loss adaptively prioritizes hard-to-generate frequency components \cite{jiang2021focalfrequencylossimage}. Some methods also try to either remove high-frequency information \cite{yamaguchi2021fdropmatchgansdeadzone} or perturb it \cite{li2021exploringeffecthighfrequencycomponents}, thereby mitigating the discriminator's tendency to exploit high-frequency features for classification.

Beyond these frequency-aware methods for GANs, other methods leverage wavelets to better supervise the generation of fine, non-stationary details \cite{korkmaz2024traininggenerativeimagesuperresolution, 8014882, 8237449, gal2021swaganstylebasedwaveletdrivengenerative}. Similar to our method, prior work used a multi-scale Laplacian pyramid approach, applying losses at different spatial resolutions to explicitly provide supervision at different frequency bands \cite{bojanowski2019optimizinglatentspacegenerative, denton2015deepgenerativeimagemodels, didwania2025laplosslaplacianpyramidbasedmultiscale}. In contrast to this prior work which directly uses the Laplacian pyramid based supervision on the full image content, we perform a frequency decomposition before our loss is computed to enforce translation of color from our source image domain to the target image domain with structural information.

\section{Method}

\begin{figure}[htbp]
  \centering
  \includegraphics[width=\linewidth]{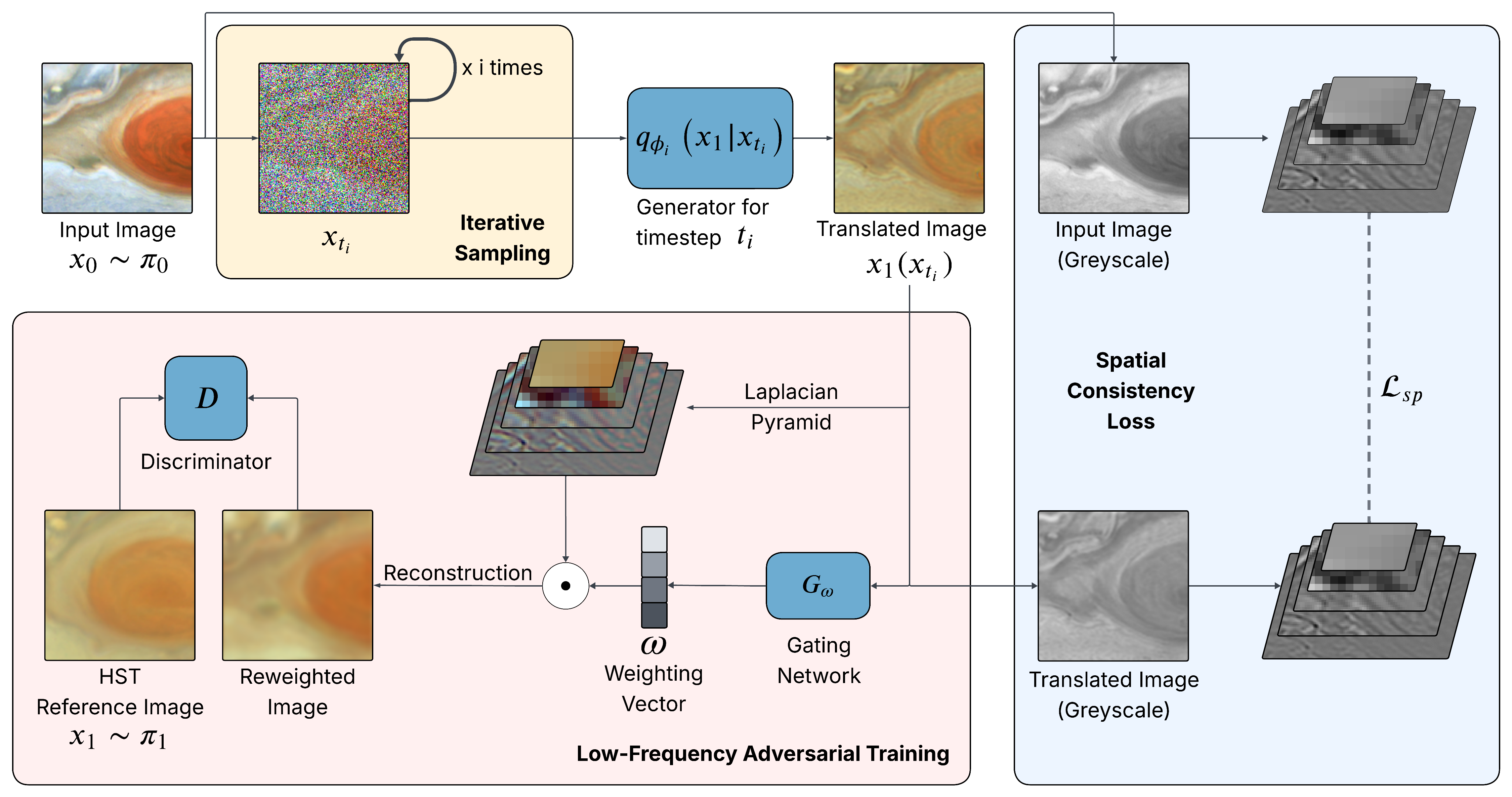}
  \caption{Our network disentangles the learning process into two streams. First, a gating network acts as a dynamic low-pass filter, reweighing the predicted image to learn low-frequency color from the HST domain. Simultaneously, a Spatial Consistency Loss explicitly retains the high-frequency spatial detail from the original JunoCam source.}
  \label{fig:method}
\end{figure}

Our primary challenge is to apply the scientifically calibrated color profile of Hubble Space Telescope (HST) observations to high-resolution JunoCam images without sacrificing JunoCam's superior spatial detail. Adversarial training in state-of-the-art unpaired image-to-image translation methods is ill-suited for this task~\cite{kim2024unpaired,cut2020}, as a discriminator trained on HST data learns to penalize the fine-scale JunoCam structures as out-of-domain artifacts.  Therefore, we introduce a new method seen in Figure~\ref{fig:method} for unpaired image-to-image translation which decouples the adversarial learning of low-frequency color from a direct consistency loss that preserves high-frequency spatial structure.

\subsection{Learning a Schrödinger Bridge}

We first formulate the unpaired image-to-image translation task as a Schrödinger Bridge Problem (SBP) following~\cite{kim2024unpaired}.
The SBP aims to find the most probable path between two distributions $\pi_0$ and $\pi_1$. Specifically, the SBP solves:
\begin{equation}
    \mathbb{Q}^{SB} = \underset{\mathbb{Q} \in \mathcal{P}(\Omega)}{\arg \min} D_{KL}(\mathbb{Q}\|\mathbb{W}^\tau) \quad \text{s.t.} \quad \mathbb{Q}_0 = \pi_0, \mathbb{Q}_1 = \pi_1 \nonumber
\end{equation}
where $\mathcal{P}(\Omega)$ is a probability measure over the space of continuous paths $\Omega$,
$\mathbb{W}^\tau$ is the Wiener measure with variance $\tau$, and $\mathbb{Q}_t$ denotes the marginal of $\mathbb{Q}$ at time $t$. $\mathbb{Q}^{SB}$ is then the Schrödinger Bridge between distributions $\pi_0$ and $\pi_1$.

We learn a Schrödinger bridge $\mathbb{Q}^{SB}$ between $\pi_0$, the uncalibrated high frequency images, and $\pi_1$, the calibrated low frequency images. We discretize $\mathbb{Q}^{SB}$ into an N step Markov Chain learnt via a composition of generators $q_{\phi_i}(x_1|x_{t_i}), i \in \{0, 1, \ldots N\}$. Each generator in this composition is trained via an adversarial constrained minimization problem:
\begin{equation}
    \min_{\phi_i} \mathcal{L} = \mathcal{L}_{\text{Adv}}(\phi_i, t_i) + \lambda_{sp}\mathcal{L}_{\text{sp}}(\phi_i, t_i) + \lambda_{SB}\mathcal{L}_{\text{SB}}(\phi_i, t_i) \nonumber
\end{equation}
where
\begin{align}
\label{eq:sb}
    \mathcal{L}_{SB}(\phi_i, t_i) &:= \mathbb{E}_{q_{\phi_i}(x_{t_i}, x_1)} [\|x_{t_i} - x_1\|^2] \nonumber\\
    & \qquad - 2\tau (1 - t_i) H(q_{\phi_i}(x_{t_i}, x_1))
\end{align}
and $\mathcal{L}_{\text{sp}}$ and $\mathcal{L}_{\text{Adv}}$ are defined in Section~\ref{sec:learnable_gate}. In Equation~\ref{eq:sb}, $x_1(x_{t_i})$ is the prediction of the generator given a state in the Markov chain, $x_{t_i}$. In practice, the sequence of generators is parametrized as a single time conditioned generation. 

\textbf{Iterative Sampling.} We iteratively sample $x_{t_{i}}$ during training via linear interpolation between $x_{t_{i-1}}$ and $x_1(x_{t_{i-1}})$, where $\{t_{i}\}_{i=0}^{N}$ is a partition on the unit interval [0, 1] and $x_1(x_{t_{i-1}})$ is the estimated target data sample given $x_{t_{i-1}}$. The sample $x_{t_{i}}$ is then passed through $q_\phi(x_1|x_{t_{i}} )$
to obtain $x_1(x_{t_i})$. The pairs $(x_{t_i} , x_1(x_{t_i} ))$
, $(x_1, x_1(x_{t_i} ))$ and $(x_0, x_1(x_{t_i} ))$ are then used to compute each component loss: $L_{SB}$, $L_{Adv}$, and $L_{sp}$. 

\subsection{Decoupling Spectral and Spatial Training with a Learnable Laplacian Gate}\label{sec:learnable_gate}

At each step $t_i$ of the generative process, the generator $q_{\phi_i}(x_1|x_{t_i})$ takes the current state $x_{t_i}$ as input to produce a prediction of the final clean image, $x_1(x_{t_i})$. This prediction is then decomposed into a standard, multi-level Laplacian pyramid, $P_{x_1(x_{t_i})}$. A secondary gate network $G_w$, acts as a content- and timestep-aware filter, taking the prediction $x_1(x_{t_i})$ and the time index $t_i$ as input, and outputting a weight vector $w = G_w(x_1(x_{t_i}), t_i)$. Each element $w_j \in [0, 1]$ corresponds to a level $j$ of the pyramid. 

These learned weights, $w$, are used to split the training objective into two distinct losses: a low frequency adversarial loss and a spatial consistency loss. By training in this manner, the gate network learns to dynamically partition the image, routing low-frequency color and style information to the discriminator while enforcing a strict, color-independent structural consistency with the original source image. We demonstrate the effectiveness of this disentanglement strategy in Appendix \ref{sec:gating_ablation}.

\textbf{Low-Frequency Adversarial Training}
We define the low-frequency component of the prediction, $x_1(x_{t_i})^L$, as the reconstruction of the pyramid re-weighted by $w$: $x_1(x_{t_i})^L = \mathcal{R}(w \odot P_{x_1(x_{t_i})})$, where $\mathcal{R}$ is the reconstruction function and $\odot$ denotes element-wise multiplication. Crucially, only $x_1(x_{t_i})^L$ and $x_1$ are passed to our discriminator $D$ during training. This forces the discriminator to base its decisions solely on the frequency components that the Gate deems relevant to the target domain, preventing it from penalizing the generator for preserving desirable source structures. The trained low-frequency discriminator is then used to produce our adversarial loss,
\begin{equation}
    \mathcal{L}_{Adv}(\phi_i, t_i) := D_{KL}(q_{\phi_i}(x_1)^L \| p(x_1)) = 0 \label{eq:sb_adv}
\end{equation}
where $p(x_1)$ is the density of the probability distribution $\pi_1$.

\textbf{Spatial Consistency Loss} To ensure high-fidelity preservation of spatial details, we enforce similarity between the prediction pyramid $P_{x_1(x_{t_i})}$ and the source image pyramid $P_{x_0}$ using a grayscale pyramid loss, ensuring sensitivity to structural differences but invariance to color: \begin{equation}      
\mathcal{L}{\text{sp}} = \sum_{j=0}^{N-1} \lambda_j \left| L_j(\mathcal{G}(P_{x_1(x_{t_i})})) - L_j(\mathcal{G}(P_{x_0})) \right|_1 
\end{equation}
where $L_j$ is the j-th pyramid level, $\mathcal{G}$ is the RGB-to-grayscale conversion, and $\lambda_j$ are fixed weights that prioritize the fidelity of high-frequency levels.

\section{Creating an Unpaired Image to Image Translation Dataset for HST and JunoCam}\label{sec:preprocess}

We construct a new dataset, JunoCam2HST, from two unpaired image datasets: high-resolution, uncalibrated JunoCam images as the source domain and lower-resolution, calibrated HST images as the target domain.

\textbf{JunoCam (Source Domain):} The source domain is constructed from raw JunoCam image strips captured during each perijove (the point of closest orbital approach to Jupiter). Initially, the 8-bit companded data is decompanded to its original 12-bit dynamic range via a lossless decompanding process~\cite{Mukai2017Juno}. A critical pre-processing step is correcting for spacecraft jitter. We perform a temporal alignment by sampling millisecond-level time offsets, selecting the value that minimizes the Euclidean distance between the observed planetary limb and its projected position. Following this correction, the aligned frames are projected onto a unified ``mid-time'' plane to establish a consistent viewpoint and then mapped to a cylindrical coordinate system. To normalize for large-scale illumination and curvature variations, we apply a ``flattening'' procedure by applying a Gaussian high pass filter, creating a clean, flattened mosaic. 

\textbf{Hubble OPAL (Target Domain):} The target domain is built from the Hubble Outer Planet Atmospheric Legacy (OPAL) program\footnote{\url{https://archive.stsci.edu/hlsp/opal}}~\cite{wong2025hubble}. We use the program's calibrated, map-projected mosaics which provide scientific reflectance data (I/F)~\cite{opal1,opal2}. These mosaics are created by stacking multiple filters to form a multi-wavelength image. 

\textbf{JunoCam2HST:} Our newly constructed dataset, JunoCam2HST, comprises 9421 JunoCam images and 5936 HST images spanning the 10 latitudinal zones of Jupiter. The two data sources are fundamentally unpaired, with no direct spatial or temporal correspondence. To create the dataset, we first generated large, 16000x16000 cylindrically-projected tiles (resolution of $\approx$ \SI{62}{km/pixel}) from the processed JunoCam mosaics and the HST OPAL maps. From these tiles, we sampled the 256x256 pixel crops used for training and evaluation.

For training, we use data from JunoCam perijoves 16, 17, 19, 20, 21, and 22 and data from HST Cycle 26 to ensure high temporal overlap. We hold out data from JunoCam perijove 18 for evaluation. The JunoCam2HST dataset will be made publicly available upon publication of this work.

\section{JunoCam Photometric Calibration Results}

\subsection{Evaluation Metrics}
A direct quantitative evaluation of our generated images is challenging due to the lack of spatially and temporally paired data between JunoCam and HST. We therefore adopt a framework inspired by multispectral pansharpening, independently evaluating the two key aspects of our translation: spatial fidelity and spectral fidelity.

\textbf{Spatial Fidelity:} To measure the preservation of fine-scale source structures, we compute the Structural Similarity Index (SSIM) and Peak Signal-to-Noise Ratio (PSNR). Both metrics are calculated between the generated output images and the corresponding original high-resolution JunoCam source images. These metrics directly assess how much spatial detail is retained after the translation.

\begin{table}[htbp]
  \centering
  \caption{Quantitative comparison on the JunoCam2HST test set. Spatial metrics (SSIM, PSNR) are computed against the source JunoCam image. Spectral ($\chi^2$) metrics are computed against the target HST domain. Our method achieves superior spatial fidelity while demonstrating a strong spectral alignment.}
  \label{tab:results}
  \sisetup{detect-weight=true, detect-family=true}
  \begin{tabular}{@{}l S[table-format=1.4] S[table-format=1.4] S[table-format=2.3]@{}}
    \toprule
    Method & {$\chi^2 \downarrow$} & {SSIM $\uparrow$} & {PSNR $\uparrow$} \\
    \midrule
    UNSB & 0.0704 & 0.7450 & 19.7700 \\
    CUT & \textbf{0.0430} & 0.7092 & 19.2751 \\

    \midrule 
     
    \multicolumn{4}{l}{\itshape \methodname} \\ 
    $L=1$ & 0.5220 & 0.8969 & 15.0811  \\
    $L=2$ & 0.1708 & 0.9451 & 24.3433 \\
    $L=4$ & 0.1677& 0.9348 & 25.8970 \\
    $L=6$ & 0.0681 & 0.9453 & \textbf{26.5560} \\
    $L=8$ & 0.0753 & \textbf{0.9454} & 23.2855 \\
     
    \bottomrule
  \end{tabular}
\end{table}

\textbf{Spectral Fidelity:} To evaluate the accuracy of the translated photometric profile, we compute the $\chi^{2}$ score. This metric measures the distance between the aggregate color histograms of the entire generated image set and the target HST image set, providing a global measure of color distribution alignment.

\subsection{Results}

\begin{figure}[htbp]
  \centering 
  \includegraphics[width=\linewidth]{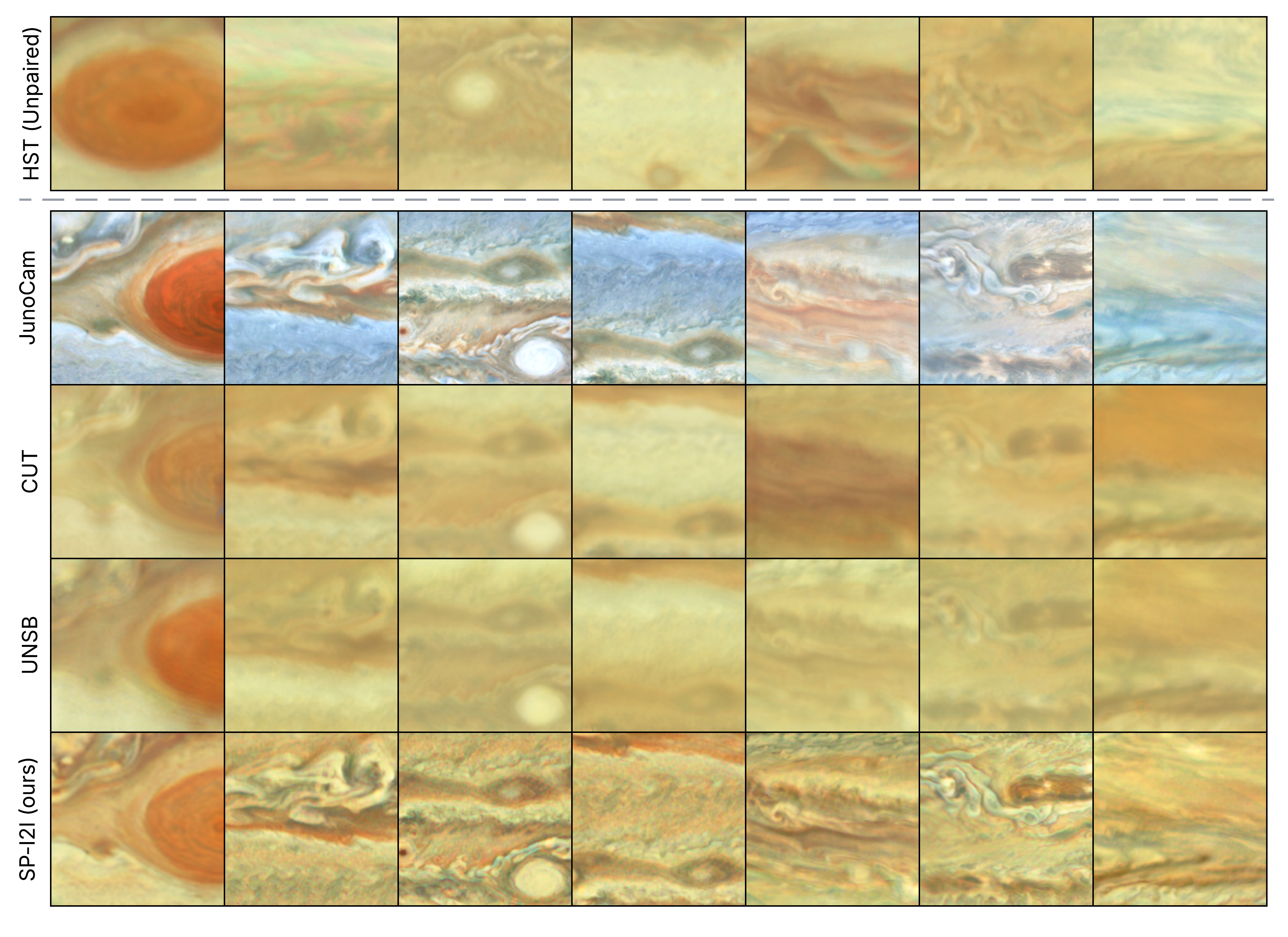}
  \caption{Qualitative comparison of our method (\methodname) against baselines. The JunoCam (Source) images exhibit rich, high-frequency detail but are uncalibrated, while the HST (Target) images show the desired calibrated color but lack this detail. Note that these images are spatially and temporally unpaired but are sampled from the same latitudinal zones. Baseline methods (CUT, UNSB) adopt the target color but fail to preserve these spatial details, resulting in blurry outputs. In contrast, \methodname successfully retains the sharp source structures while fusing them with the calibrated HST color profile.}
  \label{fig:juno_qual}
\end{figure}

We compare our method, \methodname, for structure-preserving photometric calibration of JunoCam data on our new JunoCam2HST dataset to two state-of-the-art unpaired image-to-image translation methods, UNSB~\cite{kim2024unpaired} and CUT~\cite{cut2020}, in Table~\ref{tab:results}. Our proposed method, \methodname, achieves increased spectral fidelity with a higher SSIM score and a higher PSNR \cite{psnr_ssim} value with respect to the original JunoCam image than the baselines, demonstrating better retention of high frequency JunoCam image features in the calibration process. Our method achieves a better $\chi^2$ score than UNSB but a slightly worse $\chi^2$ score than CUT. Since $\chi^2$ is compared against the HST images which contain mostly low-frequency information, we expect the baselines to perform well.
The value of the JunoCam data is largely in the high frequency features. With \methodname, we maintain a competitive $\chi^2$ distance between the aggregate color histograms of the predicted distribution and the HST dataset while simultaneously preserving high frequency information in the images. This result is qualitatively visualized in Figure~\ref{fig:juno_qual}.

\textbf{Ablation on Number of Decoupling Levels.} We analyze the impact of the number of Laplacian levels (L) used for decoupling in Table~\ref{tab:results}. We observe a strong correlation between the number of levels and the resulting model performance. The L=1 model, which attempts a minimal frequency separation, fails to effectively decouple the frequency content, resulting in low spatial fidelity (PSNR $ = $ 15.0811) and poor spectral matching ($\chi^2 =$  0.5220). As L increases, performance substantially improves across the board. We find an effective frequency separation at L=6, which achieves the highest PSNR (26.5560) and the lowest $\chi^2$ score (0.0681), indicating the best combination of spatial preservation and spectral alignment. 

\begin{figure}[htb!]
  \centering 
  \includegraphics[width=0.75\linewidth]{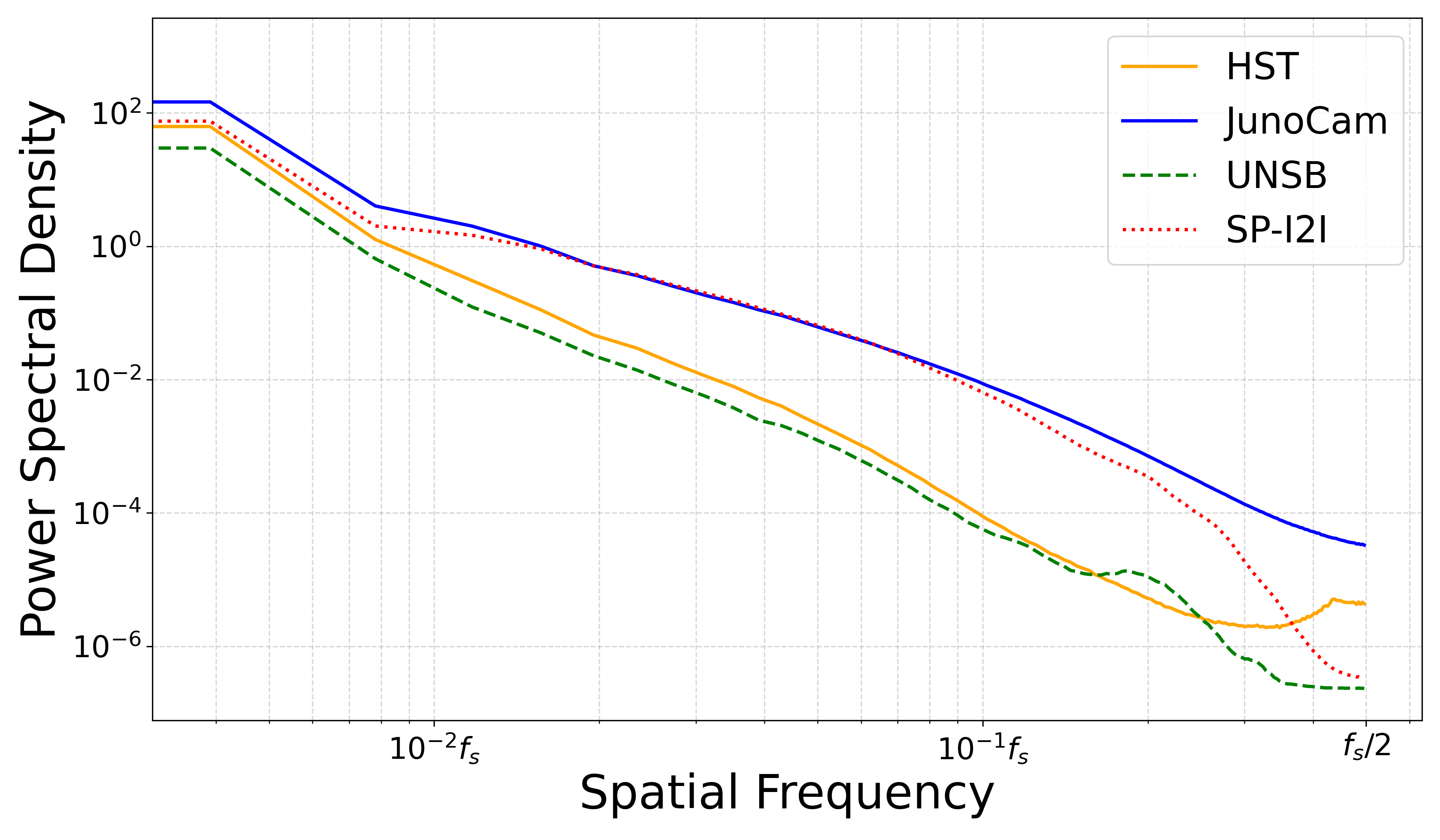}
    \caption{Average Power Spectral Density (PSD) comparison. While the baseline UNSB (green) mimics the low-detail HST target (orange), our method, \methodname  (red), uniquely matches the HST low-frequency profile while also preserving the high-frequency detail from the JunoCam source (blue). Note that $f_s/2$ is the Nyquist frequency for images.}
  \label{fig:quant_analysis} 
\end{figure}

\textbf{Frequency Content Retention.} The JunoCam imagery contains valuable structural details which we quantitatively associate with high frequency image content. We analyze the average distribution of normalized energy across frequency bands in Figure \ref{fig:quant_analysis} over our JunoCam data, HST data, and the estimated photometric calibration results from our best performing baseline, UNSB, and our proposed method, \methodname. The source JunoCam data shows significantly higher power in the mid-to-high frequency bands than the HST target, indicating a greater presence of fine-grained detail. The baseline UNSB model (green) output closely mimics the spectral profile of the HST target, effectively discarding this high-frequency information. In contrast, our proposed \methodname method (red) successfully preserves large-scale features by matching the low-frequency power, while also retaining significantly more high-frequency power from the source JunoCam images, demonstrating its ability to bridge the domain gap without sacrificing fine-grained textural detail. 

\section{Broader Impacts}
\label{sec:broad}
The number of applications of machine learning-based methods to space science studies is currently increasing, especially in subfields with more uniform/normalized data reduction and calibration methods (e.g., large surveys of stellar properties taken with a single telescope/instrument). In contrast, the subfield of planetary science has been slower to adopt these methods \cite{Azari2021}, due in part to the diversity of extant datasets (even for a single target), often making direct comparison difficult. These differences arise from variances in quality, spectral/spatial resolution, temporal coverage and observing cadence, viewing geometry, and general data modalities due to varied forms of collection - e.g., by ground-based observatories that contend with Earth’s atmosphere, by dedicated missions with instruments specifically tuned to address singular science questions, etc. Further complicating direct comparisons is the fact that most, if not all, Solar System bodies are temporally variable, rendering it difficult to cross-correlate features across these varied datasets. Historically, this has been manually addressed on a case-by-case basis for each science application \cite{Dahl2021,Moeckel2023,Brueshaber2025}

Our method is a critical first step to jumpstarting the number of tools available to the planetary science community that can be used to analyze and consolidate these diverse data. By utilizing these unpaired datasets and generating a new dataset informed by both observations, we have shown that meaningful scientific outputs can be obtained from neural network based approaches, thereby enhancing the science return of the Juno mission in a way that would otherwise be impossible via more traditional methods.

We expect that our model will not only be helpful for the Juno-HST datasets in this specific science case, but can also be widely applied across other planetary science datasets which suffer from similar heterogeneous data fusion problems. Indeed, these techniques will likely become vital tools as we approach the era of high-resolution massive datasets, such as those that will be created by current and future missions (e.g., Europa CLIPPER~\cite{korth2025europa}, JUICE~\cite{lorente2017esa}, Dragonfly~\cite{barnes2021science}, etc.), whose science returns can be enhanced beyond their original intent by applying our method in relation to lower-resolution data in other wavelength ranges. 

\subsection{Pansharpening}
\label{sec:pansharpening}
This technical challenge can be considered as an unpaired pansharpening problem. Pansharpening aims at fusing two images - a high resolution panchromatic (PAN) image and a low resolution multispectral image (MS) -  that are co-registered from a satellite, generating highly spatially aligned pairs. In contrast, we do not have images paired spatially, temporally, or by viewing geometry. Methods directly modeling the pansharpening problem assume the existence of co-registered PAN and MS images.
We compare our unpaired method to state-of-the-art paired methods for pansharpening on QuickBird-2, a standard dataset in the pansharpening community \cite{9844267}. 
To train our methods, we do not assume co-registration between PAN and MS images and train in a fully unpaired manner. Additionally, we produce baselines using off the shelf unpaired image translation methods on our data.

To evaluate our method in the reduced-resolution scenario, reconstruction accuracy is measured via PSNR \cite{psnr_ssim} and SSIM \cite{psnr_ssim}, while spectral consistency is evaluated using the Spectral Angle Mapper (SAM) \cite{sam} and the Error Relative Global Dimensionless Synthesis (ERGAS) \cite{ergas}. Spatial detail preservation is assessed using the Spatial Correlation Coefficient (SCC)~\cite{scc} and the Q8  index \cite{q_8}. For the full-resolution scenario, we use the Hybrid Quality with No Reference (HQNR) index \cite{hqnr}, which aggregates the spectral distortion, $D_{\lambda}$, and spatial distortion,  $D_s$, to evaluate the pansharpened output without a reference image.
Our quantitative results are presented in Table~\ref{tab:pansharpening_comparison}, and our qualitative results are shown in Figure~\ref{fig:pan_qual}. Additional results on the GF-2 dataset can be found in Appendix \ref{sec:gf2}.

On both datasets, our method outperforms all the other unpaired methods across all metrics. Visually, the color and detail in our qualitative results is also notably more realistic than the other unpaired methods. From our results on JunoCam calibration, we would expect this result, and due to a lack of ground truth calibrated data for evaluating our JunoCam results, these results on pansharpening support the overall evaluation of our model performance. However, there is a significant gap between the performance of all the unpaired methods and the paired methods. Enabling unpaired pansharpening would broadly add value to available imagery from the remote sensing community, and filling this performance gap with future methodological research would have implications beyond space science.

\begin{table}[htbp]
  \centering
  \caption{Comparison of methods on Quickbird. We separate methods that require co-registration (top) from those that do not (bottom).} 
  \label{tab:pansharpening_comparison}   
  \resizebox{\linewidth}{!}{%
  \begin{tabular}{@{}l S[table-format=1.3]
                      S[table-format=1.3]
                      S[table-format=1.3]
                      S[table-format=2.3] 
                      S[table-format=1.3]
                      S[table-format=2.3]
                      S[table-format=1.3]
                      S[table-format=2.3]
                      S[table-format=1.3]@{}}
    \toprule
    & \multicolumn{3}{c}{Full-Resolution} & \multicolumn{6}{c}{Reduced-Resolution} \\
    \cmidrule(lr){2-4} \cmidrule(lr){5-10} 
    Method & {HQNR$ \uparrow$} & {$D_S \downarrow$} & {$D_\lambda \downarrow$} & {ERGAS $\downarrow$} & {SCC $\uparrow$} & {SAM $\downarrow$} & {Q8 $\uparrow$} & {PSNR $\uparrow$} & {SSIM $\uparrow$} \\
    \midrule
     
    \multicolumn{10}{l}{\itshape With Co-registration} \\
    DCPNet      & 0.880 & 0.073 & 0.051 & 3.618 & 0.983 & \bfseries 4.420 & 0.935 & 38.079 & \bfseries 0.963 \\
      CANConv    & 0.893 & 0.070 & 0.039 & 3.740 & 0.982 & 4.554 & 0.935 & 37.795 & 0.960 \\
      PanCrafter & \bfseries 0.920 & 0.039 & 0.043 & \bfseries 3.570 & \bfseries 0.984 & 4.426 & \bfseries 0.938 & \bfseries 38.195 & \bfseries 0.963 \\
     
    \midrule 
     
    \multicolumn{10}{l}{\itshape Without Co-registration} \\
    UNSB        & 0.722 & 0.190 & 0.110 & 15.516 & 0.650 & \textbf{15.231} & 0.417 & 18.465 & 0.448 \\
    CUT         & 0.700 & 0.106 & 0.217 & 11.371 & 0.766 & 19.206 & 0.630 & 24.638 & 0.741 \\
    Ours        & \textbf{0.818} & \textbf{0.087} & \textbf{0.106} & \textbf{10.549} & \textbf{0.884} & 16.418 & \textbf{0.783} & \textbf{28.480} & \textbf{0.755} \\
     
    \bottomrule
  \end{tabular}%
  }
\end{table}

\begin{figure}[h!]
  \centering 
  \includegraphics[width=\linewidth]{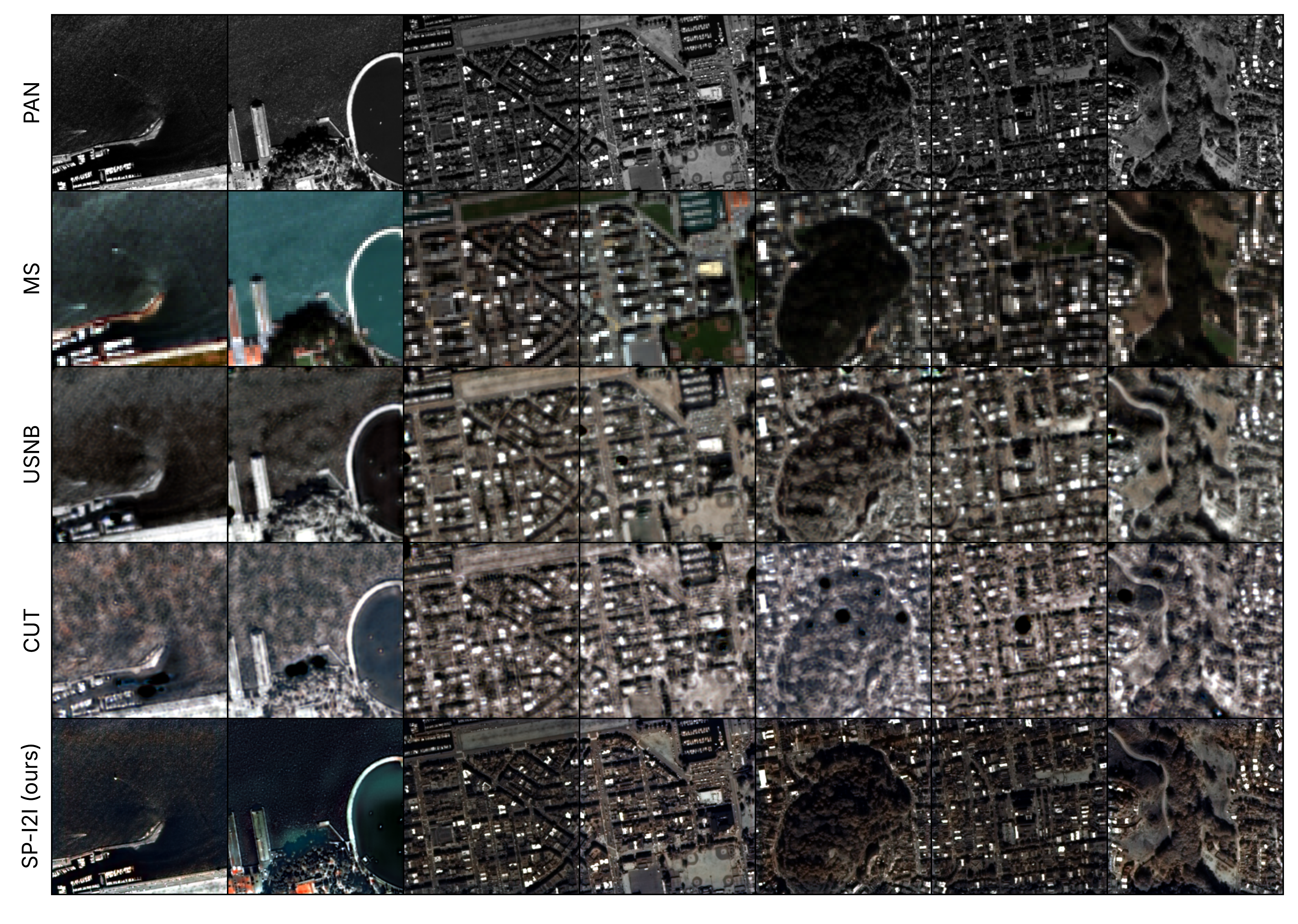}
  \caption{Qualitative comparison of our method (\methodname) against unpaired baselines on the QuickBird dataset, assuming no co-registration.}
  \label{fig:pan_qual}
\end{figure}

\section{Conclusion}

We introduced a frequency-aware approach to unpaired image-to-image translation designed to calibrate uncalibrated JunoCam images using lower-resolution, calibrated HST data. Motivated by the need to preserve distinct high-frequency content vital for studying meso-scale Jovian phenomena present only in JunoCam, our method separates the learning process, using a Laplacian weighting scheme to filter low-frequency components - used to guide the adversarial loss for photometric style transfer - while a Laplacian high frequency consistency loss explicitly preserves fine structural details from the source images. By integrating this strategy within the UNSB framework, we mitigate the detail degradation common in standard translation techniques when faced with resolution disparities. Experiments on our curated JunoCam2HST dataset validate our method's ability to retain source-specific high-frequency information, paving the way for more quantitative scientific analysis of Jupiter's atmosphere using the valuable JunoCam dataset.
   
\textbf{Limitations.} The lack of calibrated JunoCam ground truth data poses a restriction on the direct quantitative validation of calibration accuracy. While separate spectral and spatial analysis provides some insights into our models capabilities, our spectral evaluation only focuses on distributional properties and per instance spectral evaluation remains challenging. Future work will address these limitations by performing quantitative validation by comparing derived reflectance spectra of JunoCam data calibrated using our method against calibrated instruments like HST. Finally, we are in the process of exploring uncertainty estimation techniques such as HyperDiffusion \cite{erkoç2023hyperdiffusiongeneratingimplicitneural} before using the generated, calibrated data for studies on the nature of meso-scale storms on Jupiter.

\subsubsection*{Acknowledgments}
This work was supported by NASA ROSES under Award No. 23-NFDAP23\_2-0024 issued through the New Frontiers Data Analysis Program (NFDAP). 
The research was carried out at the Jet Propulsion Laboratory, California Institute of Technology, under a contract with the National Aeronautics and Space Administration (80NM0018D0004).

\clearpage
\bibliographystyle{unsrt}  
\bibliography{references} 

\newpage
\appendix
\section{Appendix}

\subsection{Impact of the Learnable Gating Network}
\label{sec:gating_ablation}

\begin{figure}[htbp] 
  \centering 
  \includegraphics[width=\linewidth]{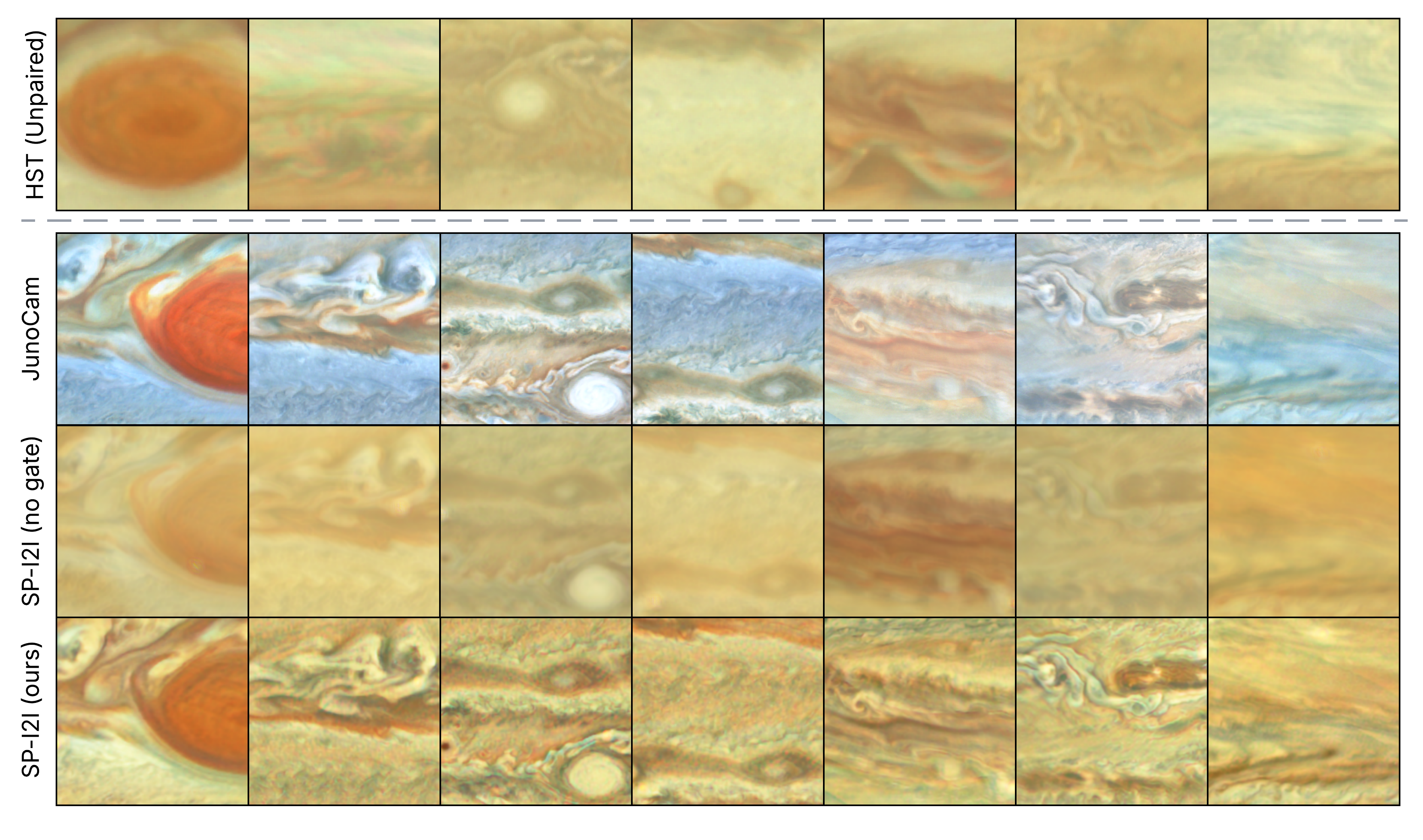}
  \caption{Qualitative ablation results. The ``no gating'' variant exhibits artifacts resulting from the optimization conflict between spectral and spatial losses, whereas our full method preserves sharp details while matching the target color profile.}
  \label{fig:gating_ab_sup} 
\end{figure}

To isolate the contribution of our decoupling strategy, we perform an ablation study where the learnable gating network is removed, effectively forcing the discriminator to evaluate the full generated image rather than filtered frequency bands. As reported in Table~\ref{tab:gating_ablation}, this "no gating" configuration significantly underperforms the full \methodname, with PSNR dropping from 26.556 to 21.825. We attribute this performance degradation to an optimization conflict between competing objectives: without the gate to filter the input, the adversarial loss penalizes high-frequency JunoCam details, viewing them as out-of-distribution artifacts relative to the low-resolution target, while the Laplacian loss simultaneously attempts to preserve them. This results in a "tug-of-war" that compromises final quality. By explicitly disentangling these tasks via the gating network, \methodname mitigates this interference, allowing the discriminator to focus on color adaptation without penalizing the structural details maintained by the spatial consistency loss. Qualitative results in Figure~\ref{fig:gating_ab_sup} visually demonstrate the degradation caused by removing the gating mechanism.

\begin{table}[htbp]
  \centering
  \caption{Ablation study analyzing the impact of the proposed gating network. Removing the gating mechanism causes the model to degrade in both spatial and spectral performance.}
  \label{tab:gating_ablation}
  \sisetup{detect-weight=true, detect-family=true}
  \begin{tabular}{@{}l S[table-format=1.4] S[table-format=1.4] S[table-format=2.3]@{}}
    \toprule
    Method & {$\chi^2 \downarrow$} & {SSIM $\uparrow$} & {PSNR $\uparrow$} \\
    \midrule
    \methodname (no gating) & 0.0887 & 0.7537 & 21.825\\ 
    \methodname (ours) & \textbf{0.0681} & \textbf{0.9453} & \textbf{26.556} \\
    \bottomrule
  \end{tabular}
\end{table}

\subsection{GF-2 Pansharpening Results}
\label{sec:gf2}

\begin{figure}[htbp] 
  \centering 
  \includegraphics[width=\linewidth]{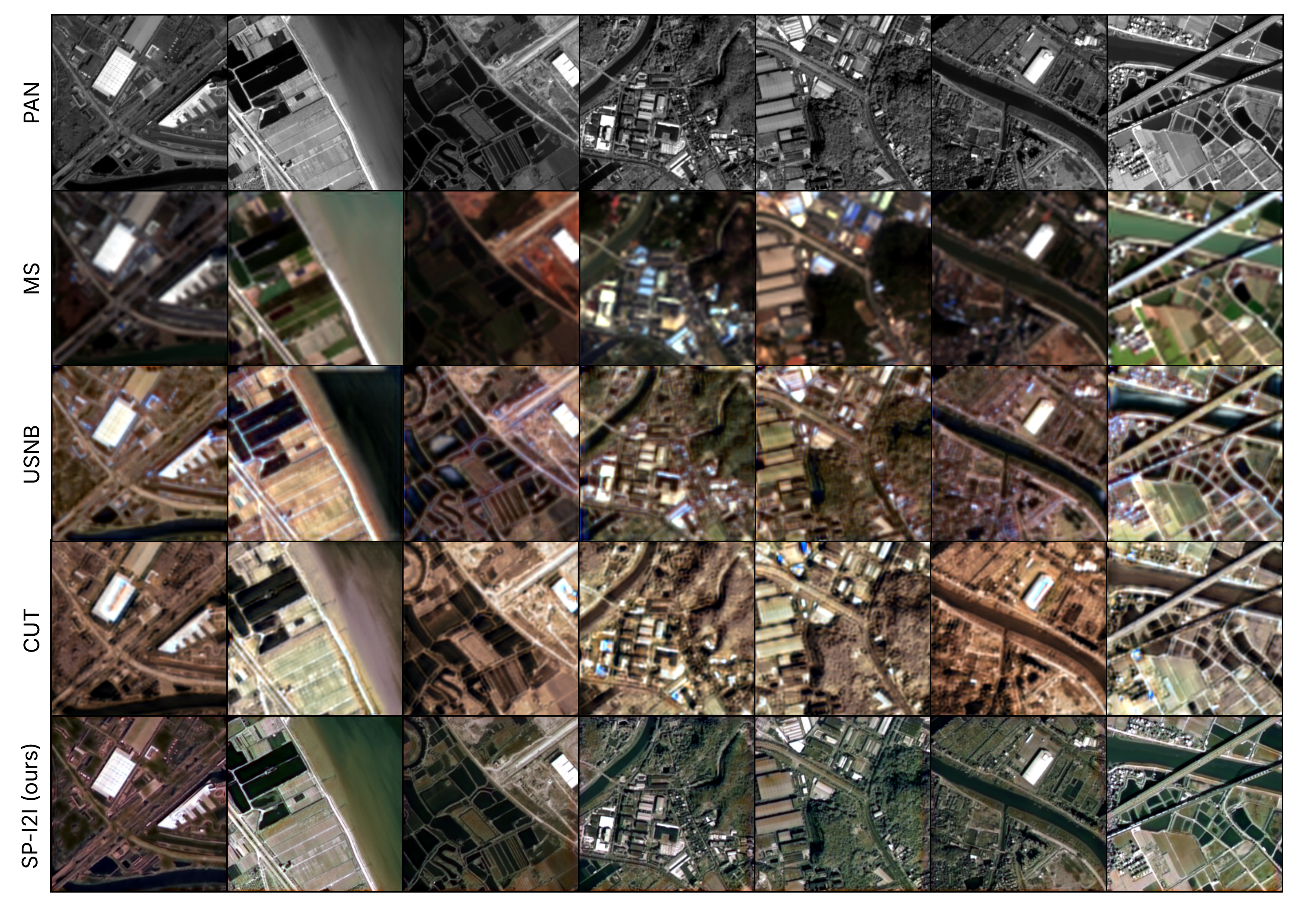}
  \caption{Qualitative results on the GF-2 dataset. Due to the strict cropping of training data to small $64 \times 64$ patches, all unpaired methods (UNSB, CUT, and Ours) lack the semantic context required to resolve spectral ambiguities (e.g., distinguishing water from vegetation). However, unlike the baselines which blur textures under these conditions, our method successfully maintains high-frequency spatial fidelity.}
  \label{fig:gf_qual} 
\end{figure}

We provide additional quantitative and qualitative results for the GF-2 \cite{9844267} dataset in Table~\ref{tab:pansharpening_comparison_gf2} and Figure~\ref{fig:gf_qual}, respectively. 

While our method quantitatively outperforms the baselines, we observe that the spectral reconstruction in the qualitative samples remains sub-optimal for all unsupervised methods. We attribute this shared limitation to the training protocol defined by the standard benchmark \cite{9844267}, which restricts training data to small crops ($64\times64$ for PAN and $16\times16$ for MS). This patch-based strategy is optimal for \textit{paired} pansharpening methods, which rely on strong pixel-wise supervision and therefore do not require broad semantic context. In contrast, unpaired methods must infer spectral mappings based on feature distributions. At this scale, the model sees only local textures without distinct object features—for instance, a smooth patch in the PAN domain could plausibly correspond to either a body of water or an open field. Lacking the broader contextual cues (e.g., coastlines) available to human observers or full-image discriminators, all unpaired models face semantic ambiguity, leading to the color inconsistencies observed across the board.

\begin{table}[htbp] 
  \centering
  \caption{Comparison of methods on GF2. We separate methods that require co-registration (top) from those that do not (bottom).} 
  \label{tab:pansharpening_comparison_gf2}   
  \resizebox{\linewidth}{!}{%
  \begin{tabular}{@{}l S[table-format=1.3]
                      S[table-format=1.3]
                      S[table-format=1.3]
                      S[table-format=1.3] 
                      S[table-format=1.3]
                      S[table-format=1.3]
                      S[table-format=1.3]
                      S[table-format=2.3]
                      S[table-format=1.3]@{}}
    \toprule
    & \multicolumn{3}{c}{Full-Resolution} & \multicolumn{6}{c}{Reduced-Resolution} \\
    \cmidrule(lr){2-4} \cmidrule(lr){5-10} 
    Method & {HQNR $\uparrow$} & {$D_S \downarrow$} & {$D_\lambda \downarrow$} & {ERGAS $\downarrow$} & {SCC $\uparrow$} & {SAM $\downarrow$} & {Q4 $\uparrow$} & {PSNR $\uparrow$} & {SSIM $\uparrow$} \\
    \midrule
     
    \multicolumn{10}{l}{\itshape With Co-registration} \\
    DCPNet      & 0.953 & 0.024 & 0.024 & 0.724 & 0.988 & 0.806 & 0.980 & 42.312 & 0.979 \\
    CANConv     & 0.919 & 0.063 & \bfseries 0.019 & 0.653 & 0.991 & 0.722 & 0.983 & 43.166 & 0.982 \\
    PAN-Crafter& \bfseries 0.964 & \bfseries 0.017 & 0.020 & \bfseries 0.552 & \bfseries 0.994 & \bfseries 0.596 & \bfseries 0.988 & \bfseries 45.076 & \bfseries 0.988 \\
     
    \midrule 
     
    \multicolumn{10}{l}{\itshape Without Co-registration} \\
    UNSB        & 0.781 & 0.130 & \textbf{0.101} & 7.577 & 0.702 & 9.223 & 0.628 & 19.533 & 0.675 \\
    CUT         & 0.771 & 0.130 & 0.113 & 6.966 & 0.734 & 7.742 & 0.612 & 20.720 & 0.647 \\
     
    Ours        & \textbf{0.783} & \textbf{0.096} & 0.135 & \textbf{5.167} & \textbf{0.886} & \textbf{7.434} & \textbf{0.781} & \textbf{25.547} & \textbf{0.777} \\
     
    \bottomrule
  \end{tabular}%
  }
\end{table}

\subsection{Comparison of Calibrated Mosaics}
\label{sec:mosaics}

\begin{figure}[htbp] 
  \centering 
  \includegraphics[width=\linewidth]{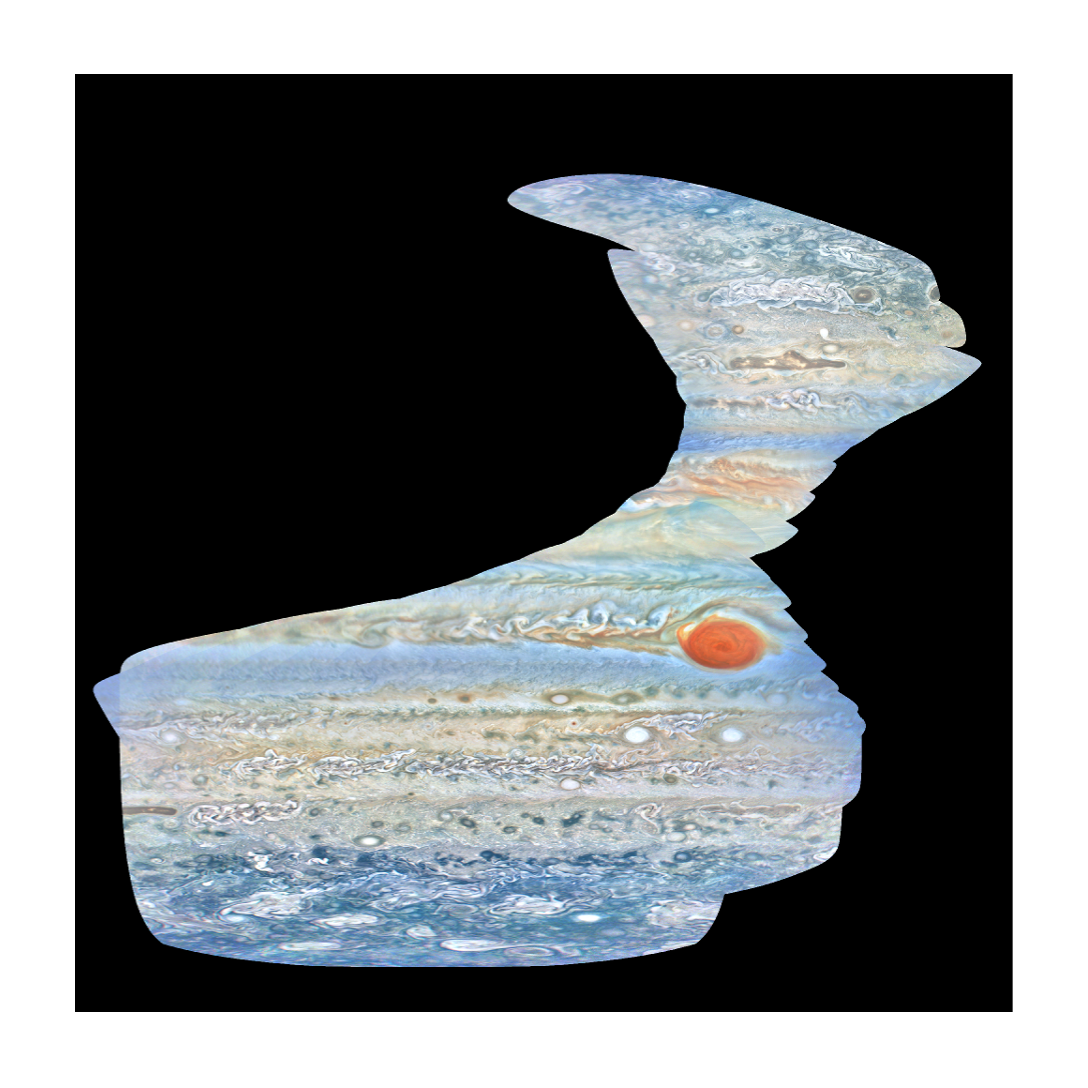}
  \caption{Visualization of the mosaic constructed using held-out test set from JunoCam Perijove 18 (PJ18).}
  \label{fig:juno_mosaic} 
\end{figure}

\begin{figure}[htbp] 
  \centering 
  \includegraphics[width=\linewidth]{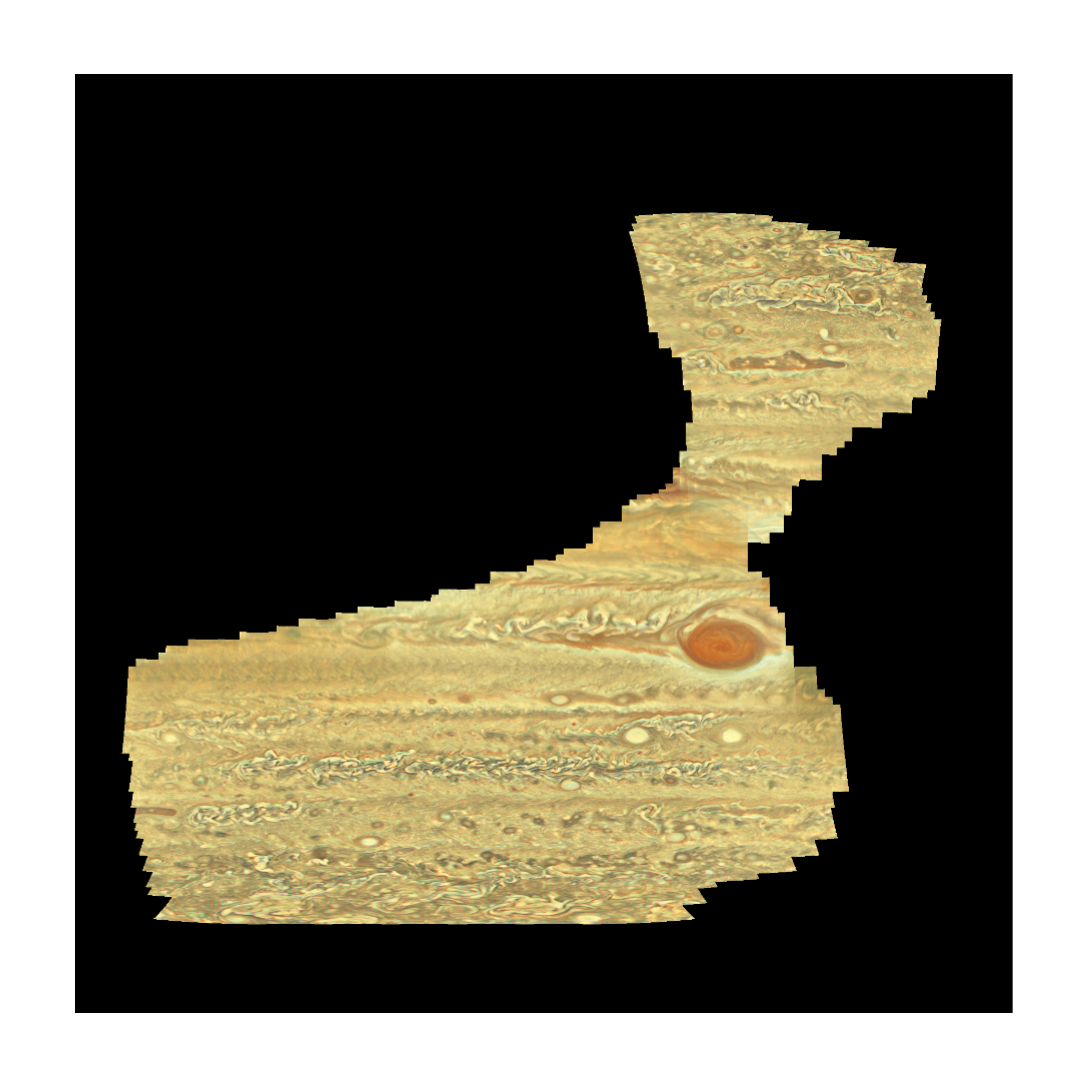}
  \caption{The calibrated output generated by \methodname on the Perijove 18 test set. Our method successfully translates the image to the target HST photometric profile while preserving the fine-scale atmospheric details visible in the original JunoCam source.}
  \label{fig:ours_mosaic} 
\end{figure}

\begin{figure}[htbp] 
  \centering 
  \includegraphics[width=\linewidth]{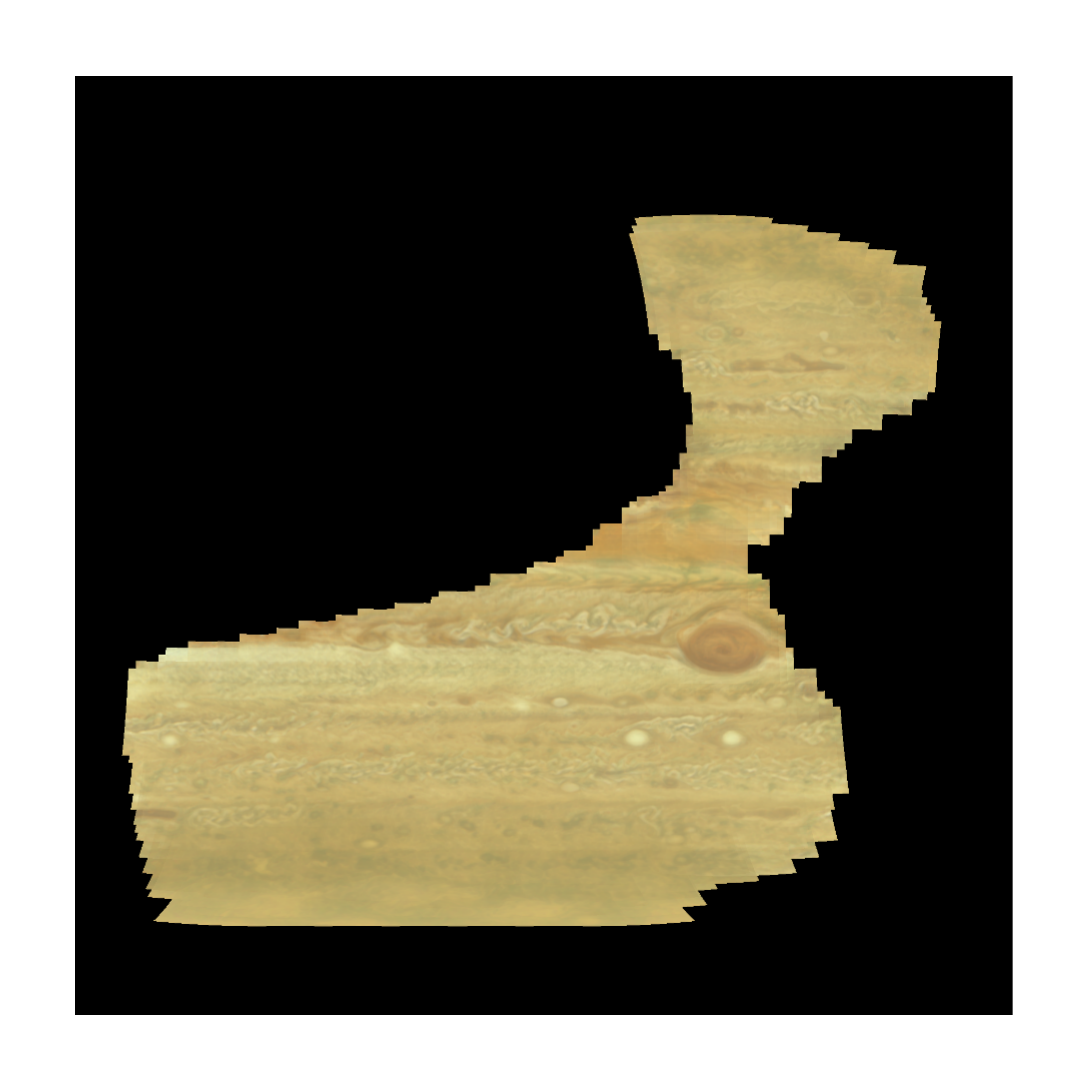}
  \caption{The output generated by the UNSB baseline on the Perijove 18 test set. While the method aligns the spectral distribution with the HST target, it fails to preserve the high-frequency spatial structures, resulting in a noticeably blurrier reconstruction compared to the source and \methodname.}
  \label{fig:unsb_mosaic} 
\end{figure}

We visually compare the calibration performance on the held-out Perijove 18 dataset \footnote{Video comparison of mosaics can be found \href{https://youtu.be/ZPtnz2bvTvE}{here.}}. These mosaics are generated by projecting individual patches to a cylindrical (equirectangular) projection and averaging the overlapping regions from multiple patches. 
The original uncalibrated JunoCam mosaic, shown in Figure~\ref{fig:juno_mosaic}, exhibits rich high-frequency atmospheric structures (e.g., turbulent distinct cloud bands and vortices). The baseline UNSB method (Figure~\ref{fig:unsb_mosaic}) successfully adapts the spectral profile to match the HST target but suffers from significant detail loss, resulting in a blurry reconstruction that obscures these meso-scale features. In contrast, our \methodname (Figure~\ref{fig:ours_mosaic}) generates calibrated images while faithfully preserving the fine-scale morphological details present in the source.

\end{document}